\documentclass[12pt,preprint]{aastex}

\begin{document}

\title{Analytical Solutions for Radiative Transfer: Implications for
Giant Planet Formation by Disk Instability}

\author{Alan P.~Boss}
\affil{Department of Terrestrial Magnetism, Carnegie Institution of
Washington, 5241 Broad Branch Road, NW, Washington, DC 20015-1305}
\authoremail{boss@dtm.ciw.edu}

\begin{abstract}

 The disk instability mechanism for giant planet formation
is based on the formation of clumps in a marginally-gravitationally 
unstable protoplanetary disk, which must lose thermal energy 
through a combination of convection and radiative cooling if
they are to survive and contract to become giant protoplanets.
While there is good observational support for forming at least
some giant planets by disk instability, the mechanism has 
become theoretically contentious, with different three dimensional
radiative hydrodynamics codes often yielding different results.
Rigorous code testing is required to make further progress.
Here we present two new analytical solutions for radiative transfer 
in spherical coordinates, suitable for testing the 
code employed in all of the Boss disk instability calculations.
The testing shows that the Boss code radiative transfer routines
do an excellent job of relaxing to and maintaining the analytical 
results for the radial temperature and radiative flux profiles 
for a spherical cloud with high or moderate optical depths,
including the transition from optically thick to optically thin 
regions. These radial test results are independent of whether the Eddington 
approximation, diffusion approximation, or flux-limited diffusion 
approximation routines are employed. The Boss code does an equally
excellent job of relaxing to and maintaining the analytical results 
for the vertical ($\theta$) temperature and radiative flux profiles 
for a disk with a height proportional to the radial distance. 
These tests strongly support the disk instability mechanism 
for forming giant planets.

\end{abstract}


\section{Introduction}

 High precision Doppler surveys provide the best
estimates to date of the frequency of gas giant planets around
solar-type stars in the sun's neighborhood of the galaxy.
Cumming et al. (2008) analyzed 8 years of HIRES spectrometer
data from the Keck Planet Search effort, begun in 1996,
taking into account the detection threshold for each star.
Cumming et al. (2008) estimated that 17\% to 20\% of
solar-type stars have planets with masses of 0.3 $M_J$
(i.e., Saturn-mass) and above, orbiting within 20 AU.
Considering that this estimate is based on an extrapolation
to longer period orbits from detections of planets with
orbital periods of 2000 days or less, this frequency estimate
may well be a lower bound, if the frequency of gas giant planets
continues to increase with orbital period: the frequency
of gas giant planets with orbital periods less than 300 days
is only 1/5 that of planets with orbital periods between
300 days and 2000 days. Gas giant planets are thought to form
primarily with orbital periods greater than 2000 days, so
there may well be an as yet unseen extensive population
of gas giant planets with orbital periods greater than 2000 days
that have not suffered inward migration. Such planets would
have formed in regions with relatively short-lived outer disks 
(e.g., Orion, Carina), while planets formed in regions with 
relatively long-lived outer disks (e.g., Taurus, Ophiuchus)
would have migrated inward (Boss 2005; Eisner et al. 2008).

 Given this high frequency of gas giant planets, it is clear 
that there must be at least one efficient mechanism for gas
giant planet formation. Two such mechanisms have been proposed,
core accretion (e.g., Mizuno 1980) and disk instability 
(e.g., Boss 1997), and both appear to be
necessary to explain the entire range of planets detected to date. 
Core accretion is the preferred mechanism for giant planets with very 
large inferred core masses, such as HD 149026b, which appears 
to have a core mass of $\sim$ 70 $M_\oplus$ and a gaseous envelope 
of $\sim$ 40 $M_\oplus$ (Sato et al. 2005). On the other hand,
disk instability would seem to be the preferred mechanism for forming 
gas giants in very low metallicity systems, such as the M4 pulsar 
planet, where the metallicity [Fe/H] = -1.5 (Sigurdsson et al. 2003).
Disk instability is also able to form gas giants around M dwarf 
stars (Boss 2006b), unlike core accretion (Laughlin et al. 2004; 
Ida \& Lin 2005). Both Doppler and microlensing searches have 
detected gas giants in orbit around M dwarfs, most recently a 
3 $M_J$ planet in the OGLE-2005-BLH-071L microlensing event 
(Dong et al. 2008). Microlensing has detected a Jupiter/Saturn
analog system (Gaudi et al. 2008) around an M dwarf that is 
consistent with photoevaporative gas loss from an outer gas giant 
formed by disk instability in a region of massive star formation
(Boss 2006b,c). Doppler searches have detected numerous
hot super-Earths, accompanied by outer gas giant planets (e.g.,
HD 181433b and HD 47186b [Bouchy et al. 2008] and HD 40307bcd
[Mayor et al. 2008]), a situation similar to that of our solar 
system and consistent with the ``best of both worlds'' scenario
(Boss 1997), where rocky planets form by collisional accumulation
inside the orbits of gas giants formed by disk instability.

 In spite of this observational evidence, theoretical work on
disk instability is split between those whose studies are either
supportive (e.g., Boss 1997, 2000, 2001, 2005, 2006a,b,c, 2007, 2008; Mayer 
et al. 2002, 2004, 2007) or dismissive (e.g., Pickett et al. 2000; 
Cai et al. 2006, 2008; Boley et al. 2006, 2007a,b; Rafikov 2005, 2007; 
Stamatellos \& Whitworth 2008) of disk instability being able
to form giant planets. This divergence appears to be the result
of a variety of modeling issues (Nelson 2006; Boss 2007), such
as numerical spatial resolution, gravitational potential solver accuracy,
artificial viscosity, stellar irradiation, radiative transfer,
and spurious numerical heating. A combination of joint code test
cases (e.g., Boss 2007) and rigorous testing is needed to try
to resolve these differences.

 Boley et al. (2006, 2007b) presented the results of a
series of tests of their radiative hydrodynamics code on a
``toy problem'' (a plane-parallel grey atmosphere) with sufficient 
assumptions to permit an analytical solution for the vertical
temperature and radiative flux profiles.
This toy problem is well-suited to their cylindrical 
coordinate code, as their vertical cylindrical coordinate can be 
used to simulate a plane-parallel atmosphere. However, the
Boley et al. (2006, 2007b) test cases are unsuitable for a code
based on spherical coordinates (Boss 2008), as is the situation for 
the Boss \& Myhill (1992) code used in Boss (1997, 2000, 2001, 
2005, 2006a,b,c, 2007, 2008).

 Boley et al. (2006) ``... challenge all researchers who publish 
radiative hydrodynamics simulations to perform similar tests or 
to develop tests of their own and publish the results.'' This
purpose of this paper is to address this challenge by the latter
route, as the former route is precluded. Two analytical solutions 
for radiative transfer in spherical coordinates are first
derived and then used to test extensively the spherically symmetric
(one dimensional, $r$) and azimuthally symmetric (two dimensional,
$r$, $\theta$) versions of the Boss \& Myhill (1992) three 
dimensional code ($r$, $\theta$, $\phi$), the same code that has 
been used in all of the author's radiative hydrodynamics models 
of disk instability.

\section{Numerical Methods}

 The new test calculations were performed with either the spherically 
symmetric or axisymmetric versions of the code that solves the three 
dimensional equations of hydrodynamics and radiative transfer, as 
well as the Poisson equation for the gravitational potential. This 
code has been used in all of the author's previous studies of disk 
instability, and is second-order-accurate in both space and time 
(Boss \& Myhill 1992). 

 For the spherically symmetric tests (R models),
the equations are solved on a spherical coordinate grid with $N_r = 101$.
The radial grid ($r$) is uniformly spaced with $\Delta r = 0.20$ AU between 
0 and 20 AU. The gravitational potential of the spherical cloud is 
obtained by the tridiagonal matrix inversion technique. The boundary 
conditions are chosen at both 0 and 20 AU to absorb radial velocity 
perturbations, though we shall see that in the new test cases,
the velocities are assumed to be zero, so the hydrodynamics
boundary conditions are irrelevant for testing the radiative
transfer routines.

 For the axisymmetric tests (T models), the equations are solved on 
a spherical coordinate grid with $N_r = 101$ and $N_{\theta} = 22$ 
for $\pi/2 \ge \theta > 0$. Symmetry through the midplane is assumed 
for $\pi \ge \theta > \pi/2$. The radial grid ($r$) is uniformly 
spaced with $\Delta r = 0.16$ AU between 4 and 20 AU, the same
configuration as is typically assumed in the Boss disk instability 
models. The Poisson equation for the disk's gravitational potential 
is solved by a spherical harmonic expansion ($Y_{lm}$) including 
terms up to $N_{lm} = 16$. Because of the axisymmetric assumption,
terms with $m \ne 0$ are dropped, reducing the solution to an
expansion in Legendre polynomials. The boundary conditions are 
chosen at both 4 and 20 AU to absorb radial velocity perturbations, 
though as noted above, this is irrelevant for the radiative
transfer.

\section{Analytical Radiative Transfer Solutions}

 Boley et al. (2006, 2007b) developed test problems based on
the approach taken by Hubeny (1990) in deriving simplified
analytical models for the vertical structure of accretion 
disks. Hubeny (1990) studied one dimensional, plane-parallel
atmospheres in hydrostatic equilibrium (i.e., at rest), with
grey opacities and Eddington approximation radiative transfer,
while neglecting convection and external radiation. Boley et al. 
(2007b) made similar assumptions for testing their flux-limited 
diffusion approximation numerical models, while also assuming 
constant vertical gravity, constant opacity, and constant density. 
We will make similar approximations here.

 All of the author's disk instability models since Boss (2001) have
employed radiative transfer in the diffusion approximation, through
the solution of the equation determining the evolution of the 
specific internal energy $E$

$${\partial (\rho E) \over \partial t} + \nabla \cdot (\rho E {\bf v}) =
- p \nabla \cdot {\bf v} + \nabla \cdot \bigl[ { 4 \over 3 \kappa \rho}
\nabla ( \sigma T^4 ) \bigr], $$

\noindent
where $\rho$ is the total gas and dust mass density, $t$ is time,
${\bf v}$ is the velocity of the gas and dust (considered to be a
single fluid), $p$ is the gas pressure, $\kappa$ is the Rosseland
mean opacity, $\sigma$ is the Stefan-Boltzmann constant, and
$T$ is the gas and dust temperature. The energy equation 
is solved explicitly in conservation law form, as are the 
other hydrodynamic equations. Boss (2008) tested a flux-limiter
for the diffusion approximation but found little difference
between disk instability models with and without the flux-limiter.

 The author's diffusion approximation code is derived from a
code that handles radiation transfer in the Eddington approximation
(Boss 1984; Boss \& Myhill 1992). In the Eddington approximation
code, the energy equation is

$${\partial (\rho E) \over \partial t} + \nabla \cdot (\rho E {\bf v}) =
- p \nabla \cdot {\bf v} + L, $$

\noindent
where $L$ is the rate of change of internal energy due to 
radiative transfer. The definition of $L$ depends on
the optical depth as

$$L = 4 \pi \kappa \rho (J-B)  \ \ \ ......... \ \ \ \tau < \tau_c, $$

$$L = {4 \pi \over 3} \nabla \cdot ({1 \over \kappa \rho} \nabla J)
\ \ \  ...... \ \ \ \tau > \tau_c,  $$
 
\noindent
where $J$ is the mean intensity, $B$ is the Planck function
($B = \sigma T^4 / \pi$), $\tau$ is the optical depth, and 
$\tau_c$ is a surface value ($\tau_c \sim 1$). 
The energy equation is solved along with the mean intensity 
equation for $J$, given by

$$ {1 \over 3} {1 \over \kappa \rho} \nabla \cdot ( {1 \over \kappa \rho} 
\nabla J) - J = -B. $$

\noindent
In the diffusion approximation, $J = B$. Hence
in both the Eddington and diffusion approximations, $L$ is
effectively given by

$$ L = {4 \pi \over 3} \nabla \cdot ({1 \over \kappa \rho} \nabla J) $$

\noindent
in optically thick regions.

 In terms of the net flux vector ${\vec H}$, $L$ is given by

$$ L = - 4 \pi \nabla \cdot {\vec H} $$

\noindent
in optically thick regions (Boss 1984), and so

$$ {\vec H} = - {1 \over 3 \kappa \rho} \nabla J. $$

\noindent
Because ${\vec H} = {\vec F} / 4 \pi$, where ${\vec F}$ is the 
radiative flux, one also finds

$$ {\vec F} = - {4 \pi \over 3 \kappa \rho} \nabla J. $$

\noindent

 We seek a solution that decouples the radiative transfer 
from the hydrodynamics. As in Hubeny (1990), we assume that
the test cloud or disk is at rest (${\bf v} = 0$). With this assumption,
the energy equation becomes

$${\partial (\rho E) \over \partial t} = L + E_H, $$

\noindent
where $E_H$ is a heating term constant in time. In steady state, then, 
the heating term must equal $- L$, so that 

$$ E_H = - {4 \pi \over 3} \nabla \cdot ({1 \over \kappa \rho} \nabla J). $$

\subsection{Radial Analytical Radiative Transfer Solutions}

 Assuming the spherically symmetric test cloud has constant, uniform 
density $\rho$ and constant, uniform opacity $\kappa$, then

$$ E_H = - {4 \sigma \over 3 \kappa \rho} {1 \over r^2} 
{\partial \over \partial r} 
\biggl( r^2 {\partial T^4 \over \partial r} \biggr). $$

\noindent
The following radial temperature distribution is consistent with 
$E_H$ constant in both space (inside the disk) and time

$$T(r) = T_o \biggl( 2 - \biggl( { r \over r_o } \biggr)^2 \biggl)^{1/4}, $$

\noindent
where $T_o$ and $r_o$ are arbitrary constants. With this solution for
$T(r)$, one finds that

$$ E_H = { 8 \sigma T_o^4 \over \kappa \rho r_o^2 }. $$

\noindent
In optically thick regions, the radiative flux ${\vec F}$ is

$$ {\vec F} = - {4 \sigma \over 3 \kappa \rho} \nabla T^4. $$

\noindent
Given the above temperature profile, ${\vec F}$ is equivalent 
to a scalar field $F(r)$ in the radial direction that is given by

$$ F(r) = {8 \sigma T_o^4 \over 3 \kappa \rho r_o^2} r $$

\noindent
in optically thick regions. The radiative flux in 
optically thick regions is calculated in the numerical code using

$$ {\vec F} = - {4 \pi \over 3 \kappa \rho} \nabla J. $$

\noindent

The radius $r_o$ is defined to be the
edge of the optically thick region, where $T(r_o) = T_o$,
and outside of which $T(r) = T_o$, i.e., the cloud is
assumed to be embedded in a thermal bath at a temperature
of $T_o = 50$ K, as is commonly assumed in the Boss disk instability
models. The envelope density ($r > r_o$) is taken to be a factor of
$10^5$ times smaller than in the optically thick region ($r < r_o$),
ensuring its optical thinness. The heating term $E_H = 0$ 
for $r > r_o$, so that the radiative flux in the optically
thin region must fall off with distance as $r^2$, i.e.,

$$ F(r) = {8 \sigma T_o^4 \over 3 \kappa \rho r_o} 
\biggl( {r_o \over r} \biggr)^2 $$

\noindent
for $r > r_o$.  
Because the numerical code uses $L$ rather than ${\vec F}$ to
calculate radiative transfer, there is no convenient expression
for ${\vec F}$ in optically thin regions that can be compared
to the analytical value above. In practice, the Boss disk
instability models assume the presence of a thermal bath
in the low optical depth regions, so the inability to compare
the numerical radiative flux with the analytical value in
low optical depth regions is not a problem.

\subsection{Vertical Analytical Radiative Transfer Solutions}

 For a disk where the height is proportional to the radial distance 
and with constant, uniform density $\rho$ and constant, uniform 
opacity $\kappa$, then

$$ E_H = - {4 \sigma \over 3 \kappa \rho} {1 \over r^2 sin\theta } 
{\partial \over \partial \theta} 
\biggl( sin\theta {\partial T^4 \over \partial \theta} \biggr). $$

The following vertical (i.e., $\theta$) temperature distribution is 
consistent with $E_H$ constant in time

$$T(\theta)  = T_o (sin\theta)^{1/4}, $$

\noindent
for $\theta > \theta_o$, where $T_o$ and $\theta_o$ are arbitrary 
constants. With this solution for $T(\theta)$, one finds that
$E_H$ varies inside the disk as

$$ E_H = - { 4 \sigma T_o^4 \over 3 \kappa \rho } 
\biggl( {cos2\theta \over sin\theta} \biggr) { 1 \over r^2 }, $$

\noindent
for $\theta > \theta_o$, with $E_H = 0$ for $\theta < \theta_o$.
Note that $E_H > 0$ only for $3\pi/4 > \theta > \pi/4$, restricting
the disk thickness about the midplane to $< \pi/4$. Note that
$\theta = 0$ corresponds to the $\hat z$ axis, the rotational
and symmetry axis, whereas $\theta = \pi/2$ corresponds to
the equatorial plane (midplane) of the disk. The calculations
all assume symmetry above and below the midplane, so that only the 
time evolution of the upper half of the disk is calculated.
 
In optically thick regions, the radiative flux ${\vec F}$ is equivalent 
to a scalar field $F(\theta)$ in the vertical direction ($\hat \theta$)
that is given by

$$ F(\theta) = - {4 \sigma T_o^4 \over 3 \kappa \rho } 
{ cos\theta \over r }, $$

\noindent
with the sign chosen to correspond to the $\theta$ coordinate,
which increases {\it downward} in the disk. The radiative flux in 
optically thick regions is calculated in the numerical code using

$$ F(\theta) = - {4 \pi \over 3 \kappa \rho} { 1 \over r }
{ \partial J \over \partial \theta}. $$

\noindent
The angle $\theta_o$ is defined to be the upper edge of the 
optically thick disk, outside of which 
$T(\theta) = T_o (sin\theta_o)^{1/4}$, i.e., 
the disk is once again assumed to be embedded in a thermal bath 
at a temperature of $T_o \approx 50$ K. The envelope density 
($\theta < \theta_o$) is taken to be a factor of $10^5$ times 
smaller than in the optically thick region ($\theta > \theta_o$).

\section{Results}

 We now present the results of two set of models that test the 
Boss code against these two analytical solutions. The codes used
are identical to the radiative hydrodynamics codes used
in the Boss disk instability models, with the exceptions being 
that in order to match the assumptions of the analytical solutions, 
the codes are restricted to either spherical symmetry or axisymmetry, 
the density ($\rho = 10^{-10}$ g cm$^{-3}$)
and opacity ($\kappa = 1$ cm$^2$ g$^{-1}$) are held 
constant, and the velocities are all set equal to zero, so
that there is no advection of energy, only heating by the $E_H$ term 
and transport by the radiative transfer term $L$. All other
aspects of the Boss codes are retained, e.g., the specific
internal energy and pressure subroutines and the subroutine
that updates the temperature based on the specific internal
energy. 

\subsection{Radial Test Results}

 Model R+50 was calculated with diffusion approximation radiative
transfer. This model was started with the analytical solution
for the initial temperature profile, but with a 50\% temperature
perturbation between 4 and 5 AU, as shown in Figure 1.
Figures 1, 3, and 5 show the resulting time evolution of the 
numerical radial temperature profile for model R+50, while Figures 2, 4, 
and 6 show the corresponding evolution of the radiative flux,
compared in each case to the steady state, analytical profiles 
(solid lines). These figures show the expected behavior in
response to the initial temperature perturbation: extra radiative
energy flows both inwards and outwards (Figure 2) as the
cloud attempts to dispel the thermal perturbation. The
perturbation extends across the entire radial extent of
the cloud (Figure 3), while the radiative flux begins
to return to the steady state value (Figure 4). Within
a few Myr, the steady state solution is attained once again,
and the code holds this solution indefinitely in time
(Figures 5 and 6). 

The numerical temperature profile at 14 Myr (Figure 5)
is very close to the steady state value, but slightly 
higher in the optically thick regions inside $r_o =$ 10 AU. 
The deviations from steady state are less than 1\%. Similarly, 
the radiative flux relaxes to very close to the analytical 
values, except at the center and just inside $r_o =$ 10 AU,
where a few cells drop noticeably below the analytical 
solution. To the extent that these few numerical values differ 
from the analytical values, they err on the side of higher 
temperatures and lower radiative fluxes, i.e., they err on 
the side of discouraging cooling of the optically thick regions. 

 Models were also calculated with Eddington approximation
radiative transfer and flux-limited diffusion approximation
radiative transfer (Boss 2008), testing their ability to
maintain the analytical solution. The results for these models
were identical to those for models with the diffusion approximation,
in the latter case because the flux-limiter was never triggered by 
the radiative fluxes in the optically thick regions. In all
cases, the code maintains the steady state solution to the
same degree that is exhibited by Model R+50 in Figures 5 and 6.
Models were also calculated for a cloud with the opacity lowered
by a factor of 100, i.e., $\kappa = 10^{-2}$ cm$^2$ g$^{-1}$.
As a result, the central optical depth dropped from $\sim 10^4$
in model R+50 to $\sim 10^2$. The lower optical depth model was
calculated with the Eddington approximation code, and it maintained 
a temperature profile that was only slightly higher than 
the analytical profile throughout most of disk, again erring on 
the side of a slightly hotter disk.

 Models were also calculated with 1\% and 10\% perturbations
to the temperature, which behaved similarly to model R+50
except for their relaxation to the steady state value occurring
faster in time, as would be expected. A model was also calculated
with a negative temperature perturbation, i.e., a cloud with
the temperature reduced by 50\% from 4 to 5 AU. This cloud
also relaxed back to the steady state solution within a
few Myr, as in model R+50. Interestingly, in this model
the relaxed solution was again slightly hotter than the
steady state solution (by less than 1\%), showing that the
code relaxes to this solution independently of being approached 
from hotter or cooler temperatures than the steady state
solution. This result implies that the radiative transfer
routines, while accurate to a high degree, do result in
a slight overestimate of cloud temperatures and an underestimate
of the radiative fluxes, which errs on the side of 
discouraging disk instability's ability to produce
self-gravitating clumps.

\subsection{Vertical Test Results}

 Model T+50 was also calculated with diffusion approximation radiative
transfer (Eddington approximation radiative transfer becomes
punitively slow in multiple dimensions, because of the need
to iterate on the solution of the mean intensity equation).
Model T+50 was started with the analytical solution for the initial 
temperature profile for a disk with $\theta_o = 83.1$ degrees
(6.9 degrees above the midplane),
but with a 50\% temperature perturbation between the midplane 
and an angle of 0.7 degrees above the midplane, as shown in Figure 7.
[Note that because of the small variation with vertical height,
the vertical temperature profile appears to be almost constant when
plotted on the scale of the applied temperature perturbation.]
Figures 7, 9, and 11 show the resulting time evolution of the 
numerical vertical temperature profile for model T+50, while Figures 8, 10, 
and 12 show the corresponding evolution of the radiative flux,
compared in each case to the steady state, analytical profiles 
(solid lines). Once again, these figures show the expected behavior in
response to the initial temperature perturbation: the upward 
radiative flux increases drastically in order to cool the
midplane (Figure 8), raising the temperature of the entire
disk (Figure 9). The radiative flux then begins to return
to the steady state values (Figure 10), and by 1.9 Myr
the disk has relaxed back to the analytical solution
(Figure 11 and 12). The exception is a single grid point
at the surface of the disk, where the flux is slightly
lower than the steady state value (Figure 12), again implying 
an error on the side of slightly lower radiative cooling.

 By way of comparison, the Boley et al. (2006) code also
errs on the side of vertical disk temperatures being
hotter than in the analytical solution (their Figure 17),
though the differences are significantly larger than
occur in the present models. The Boley et al. (2007b) code,
however, does a better job of maintaining the analytical solutions.

 A model like T+50 was also run starting with a 10\% initial
temperature perturbation. Again, the model relaxed to the 
analytical solutions as shown in Figures 11 and 12, only faster.

 Figures 7 through 12 all depict the response of the disk at
a radial distance of 7.87 AU, as that is the typical distance
where clumps form in the Boss disk instability models.
The behavior of the disk relaxation process at other
radial distances is identical to that at 7.87 AU, except
that it occurs faster at smaller radii and slower at
larger radii, again as would be expected, given
the larger vertical optical depths at greater radii,
as the disk thickness increases proportionately.

\section{Conclusions}

 The primary motivation for this paper was to answer the 
challenge issued by Boley et al. (2006) for other workers
with radiative hydrodynamics codes ``... to develop tests 
of their own and publish the results.'' The new test results
show that the Boss \& Myhill (1992) radiative hydrodynamics
code does a superb job of relaxing to and maintaining  the
analytical solutions for a spherical cloud or an axisymmetric
disk derived here. This excellent performance is independent of 
whether the Eddington approximation, diffusion approximation, or
flux-limited diffusion approximation is employed for the cloud, 
as well as of the optical depth assumed. Considering that typical
disk instability calculations with the diffusion approximation
last for only $\sim 10^3$ yr, the fact that these models show
that the radiative transfer scheme is highly accurate over 
time scales of at least $\sim 10^6$ yr is reassuring
for the mechanism of giant planet formation by disk instability.

 These models presented here test only the radiative transfer
routines and other thermodynamical aspects of the Boss codes,
not the coupling between these processes and the hydrodynamics
that occurs in full disk instability models. Ideally, one
would test the full radiative hydrodynamics codes against
analytical solutions. In the absence of such solutions, one
can test the codes with respect to their ability
to represent convective motions, which do involve a coupling
of hydrodynamics and thermodynamics. Boss (2004) analyzed
in detail the convective stability of his disk instability
models, and found a good agreement between where transient,
convective-like upwellings and downwellings occurred and where the 
Schwarzschild criterion for convection was met. Boley et al. (2007b)
presented the results of several tests for convection in their
codes, finding that convection occurred when it should have
and did not occur when it should not have. Convection and
convective-like motions thus appear to be appropriately modeled 
by both the Boss and Boley et al. codes.

 Boss \& Myhill (1992) described a variety of other 
tests to which the code has been subjected, including the
standard nonisothermal test case for protostellar collapse 
(tested on two different codes by Myhill \& Boss 1993), 
whose results have since been confirmed
by Whitehouse \& Bate (2006). Further tests of the Boss \& Myhill
(1992) code have been presented as follows: spatial resolution
(Boss 2000, 2005); gravitational potential solver (Boss 2000,
2001, 2005), artificial viscosity (Boss 2006a); and radiative 
transfer (Boss 2001, 2007, 2008). Given the ongoing
theoretical debate over the viability of disk instability
for giant planet formation, it will continue to be important
for other workers to conduct their own tests of these key 
numerical issues.

\acknowledgements

 The $r$ analytical solution was derived while I was a lecturer at
the Winter School on Exoplanets at the Theoretical Institute
for Advanced Research in Astrophysics (TIARA) of the National
Tsing Hua University, in Hsinchu, Taiwan. I thank the Acting Director
of TIARA, Ronald Taam, for making possible my visit to TIARA. 
The $\theta$ analytical solution was derived in part while I was
a visitor at the Royal Observatory, Edinburgh and at St Andrews
University in Scotland. I thank Ken Rice and Ian Bonnell for making 
those visits possible, and the referee for prompting me to investigate 
this second test case as well as for other good advice. 
I also thank Sandy Keiser for computer systems 
support at DTM. This research was supported in part by NASA Planetary 
Geology and Geophysics grant NNX07AP46G, NASA Origins of Solar Systems 
grant NNG05GI10G, and is contributed in part to NASA Astrobiology 
Institute grant NCC2-1056.

\clearpage

\begin{figure}
\vspace{-2.0in}
\plotone{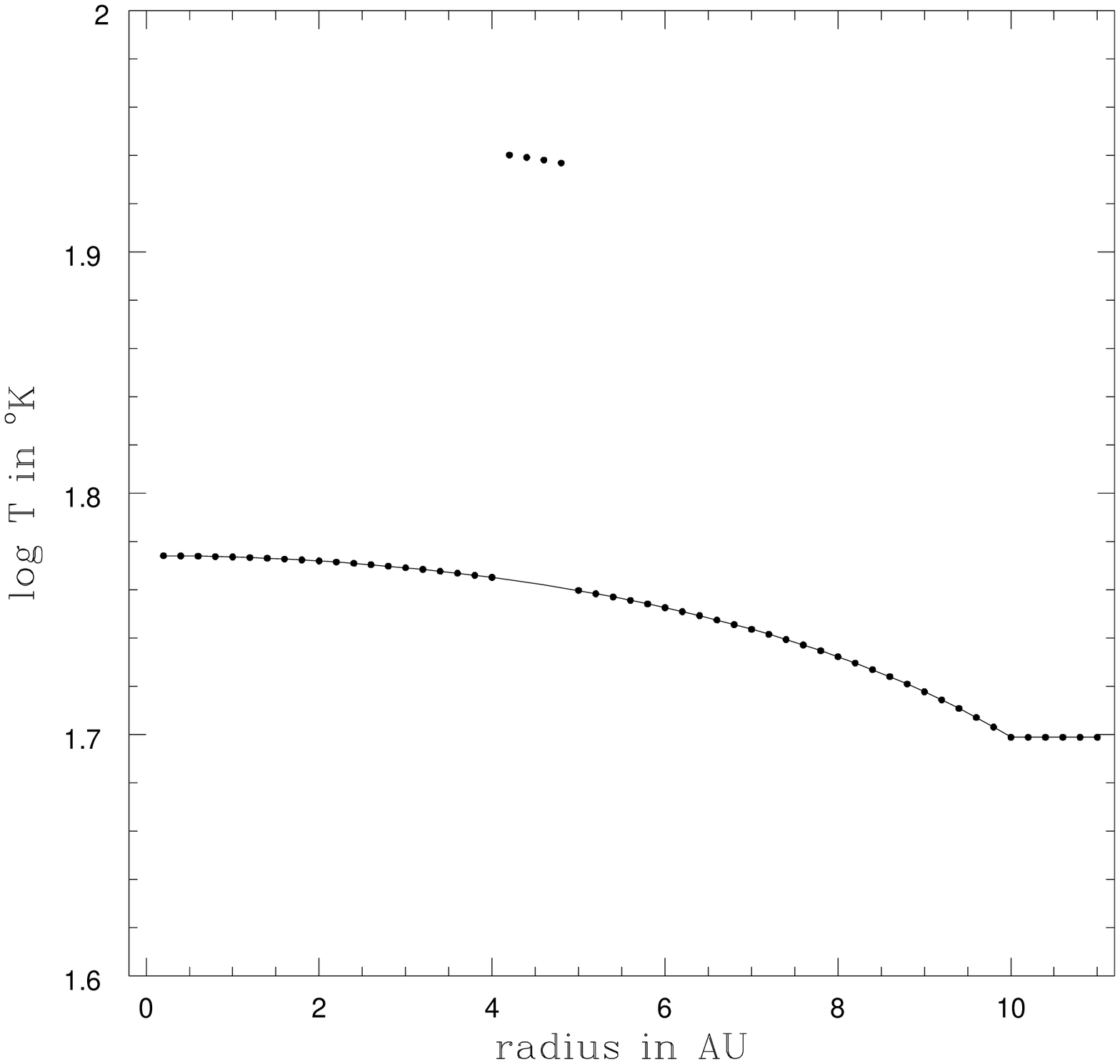}
\caption{Analytical steady-state radial temperature profile (solid line)
and numerically-calculated temperatures (solid dots) for 
model R+50 after 0.66 yr of evolution. The initial 50$\%$ 
temperature perturbation is evident between 4 and 5 AU.}
\end{figure}

\clearpage

\begin{figure}
\vspace{-2.0in}
\plotone{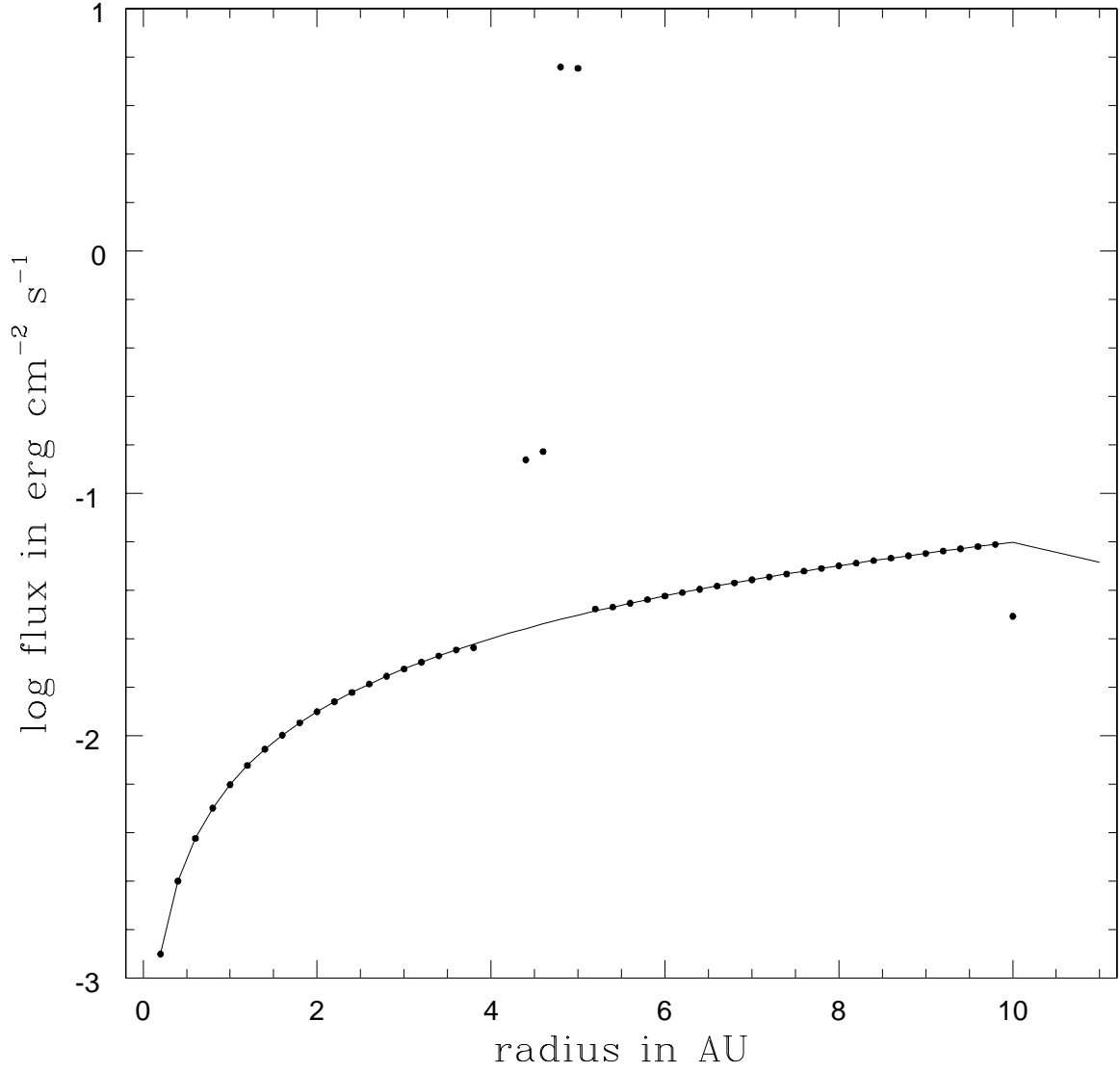}
\caption{Same as Figure 1, but for the radiative flux profile.
The numerical radiative flux is less than zero (i.e., radially inward) for 
two grid points at 4 AU and so these two points are not shown. The radiative 
flux falls to zero at the center (zero radius).}
\end{figure}

\begin{figure}
\vspace{-2.0in}
\plotone{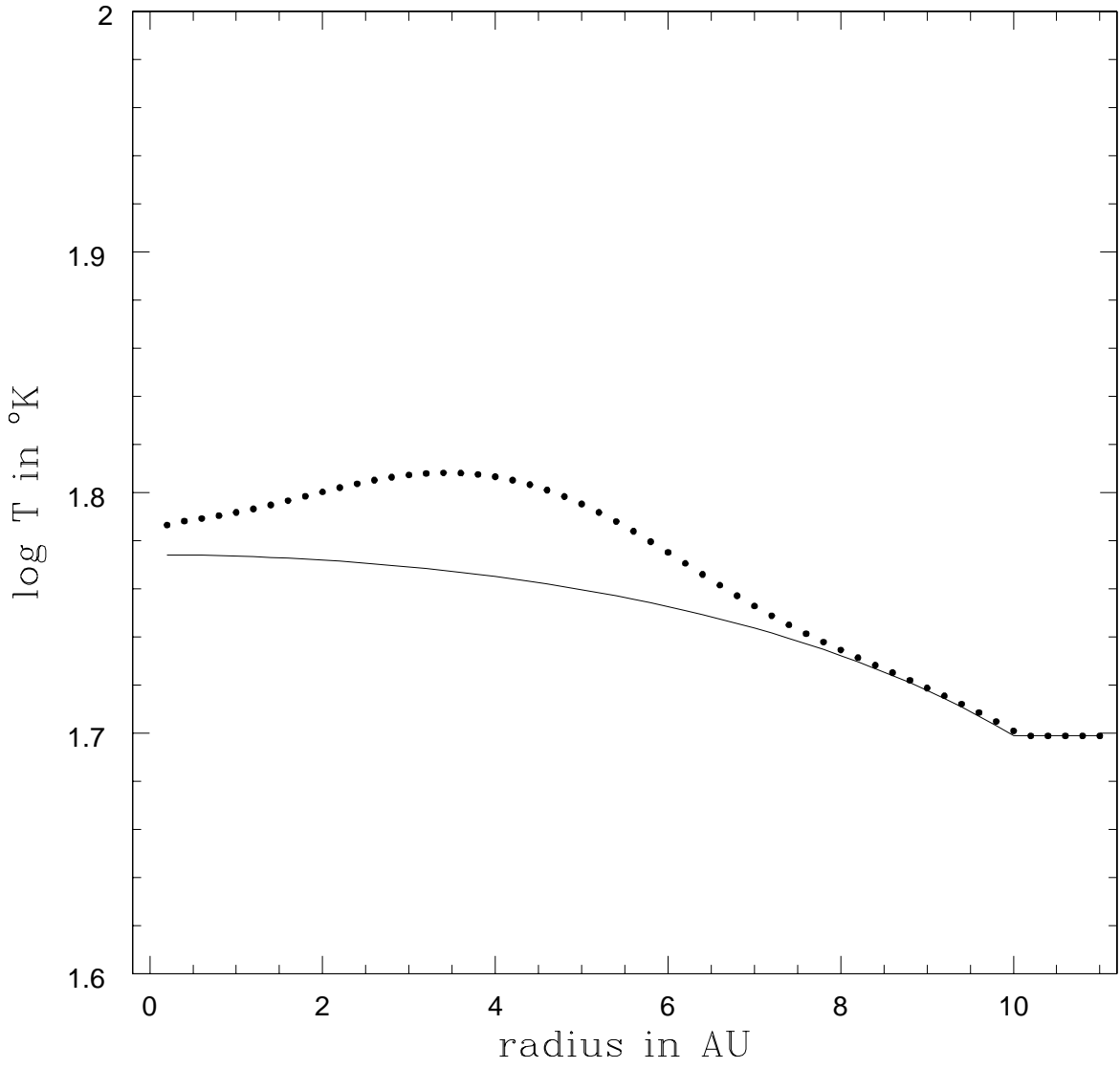}
\caption{Same as Figure 1, but after 0.071 Myr.}
\end{figure}

\clearpage

\begin{figure}
\vspace{-2.0in}
\plotone{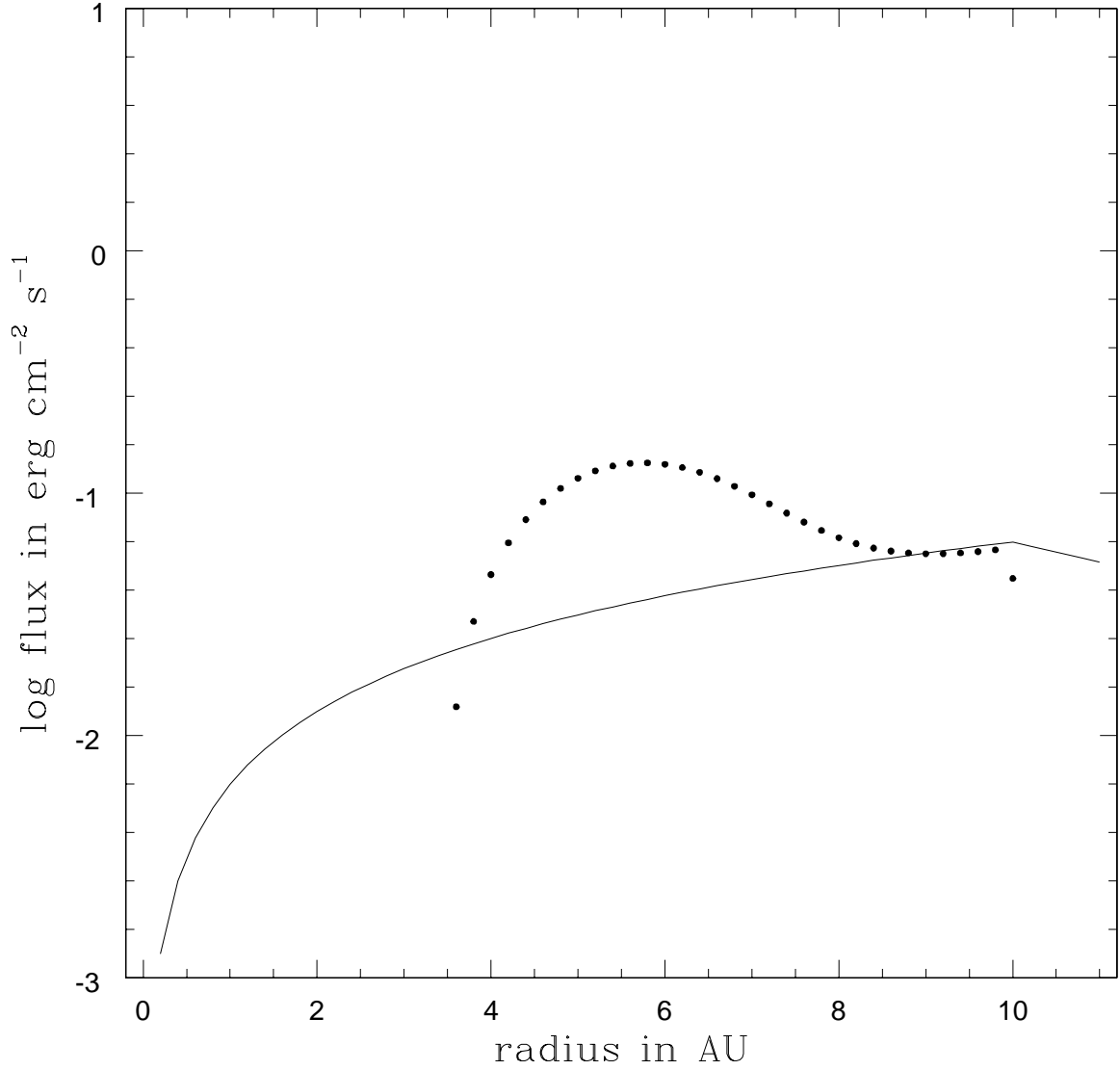}
\caption{Same as Figure 3, but for the radiative flux profile.
The radiative flux is less than zero (i.e., radially inward) for 
all the grid points interior to 3.5 AU and so these points are not shown.}
\end{figure}

\begin{figure}
\vspace{-2.0in}
\plotone{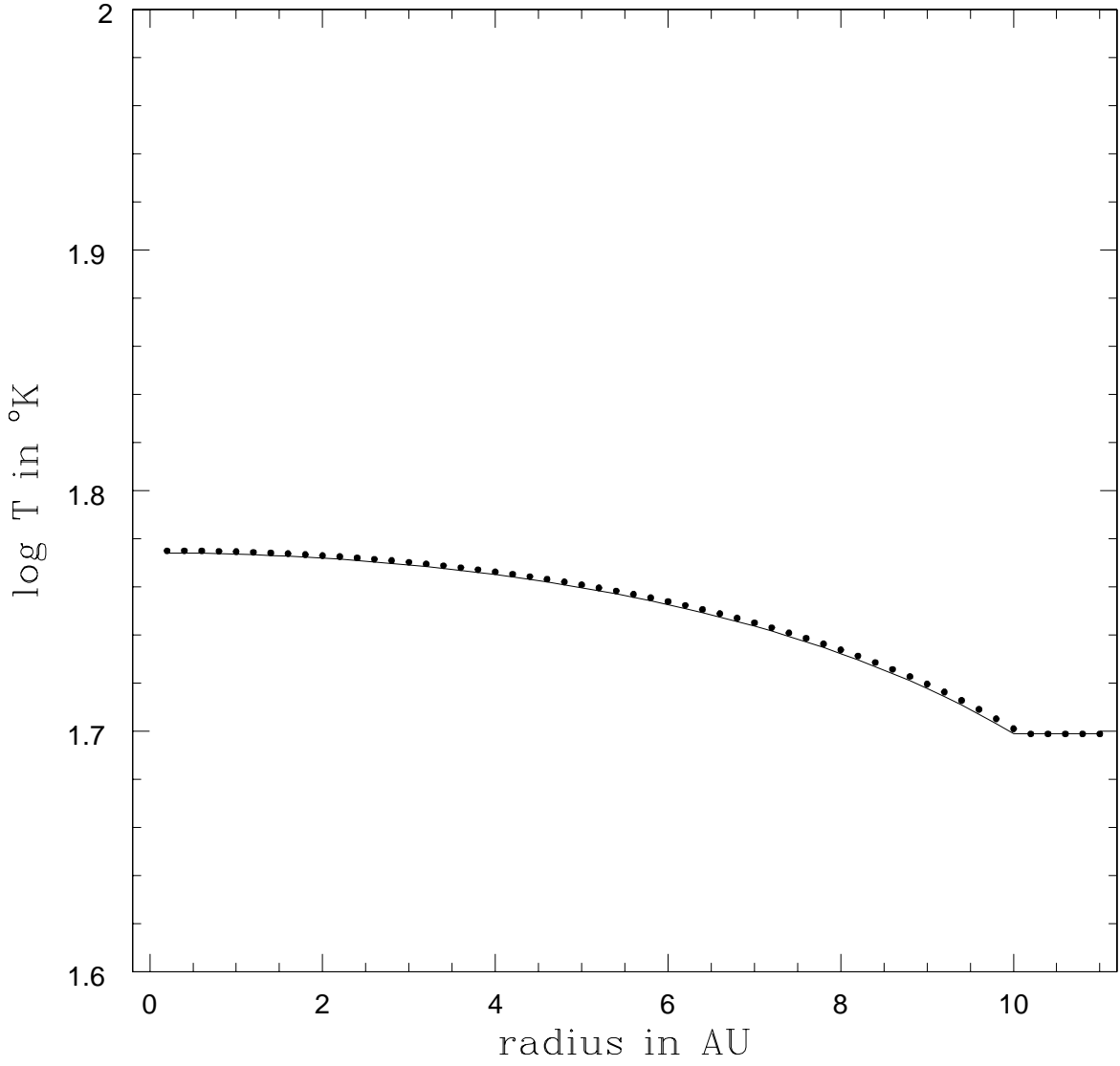}
\caption{Same as Figures 1 and 3, but after 14 Myr.}
\end{figure}

\clearpage

\begin{figure}
\vspace{-2.0in}
\plotone{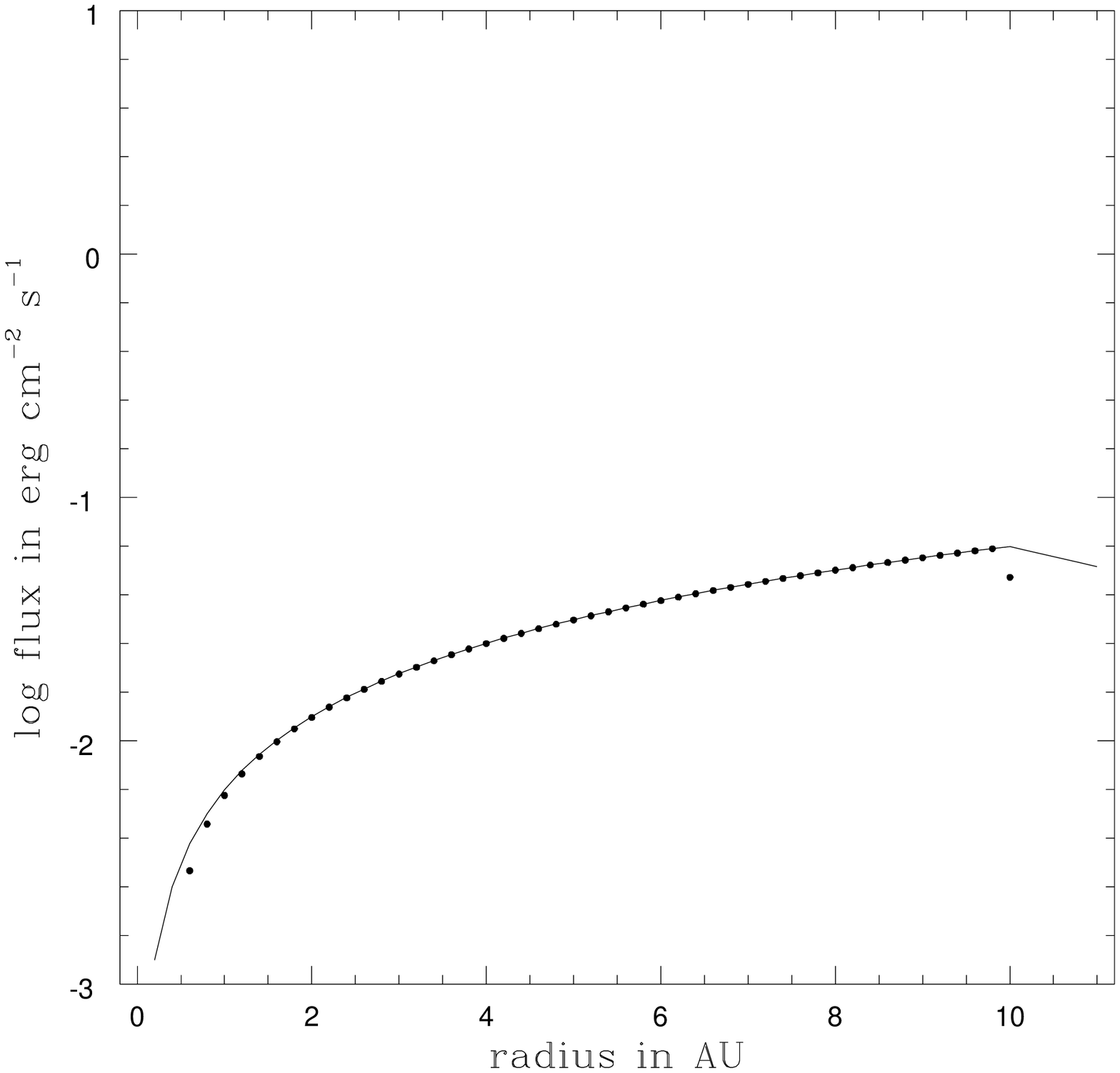}
\caption{Same as Figure 5, but for the radiative flux profile.}
\end{figure}

\begin{figure}
\vspace{-2.0in}
\plotone{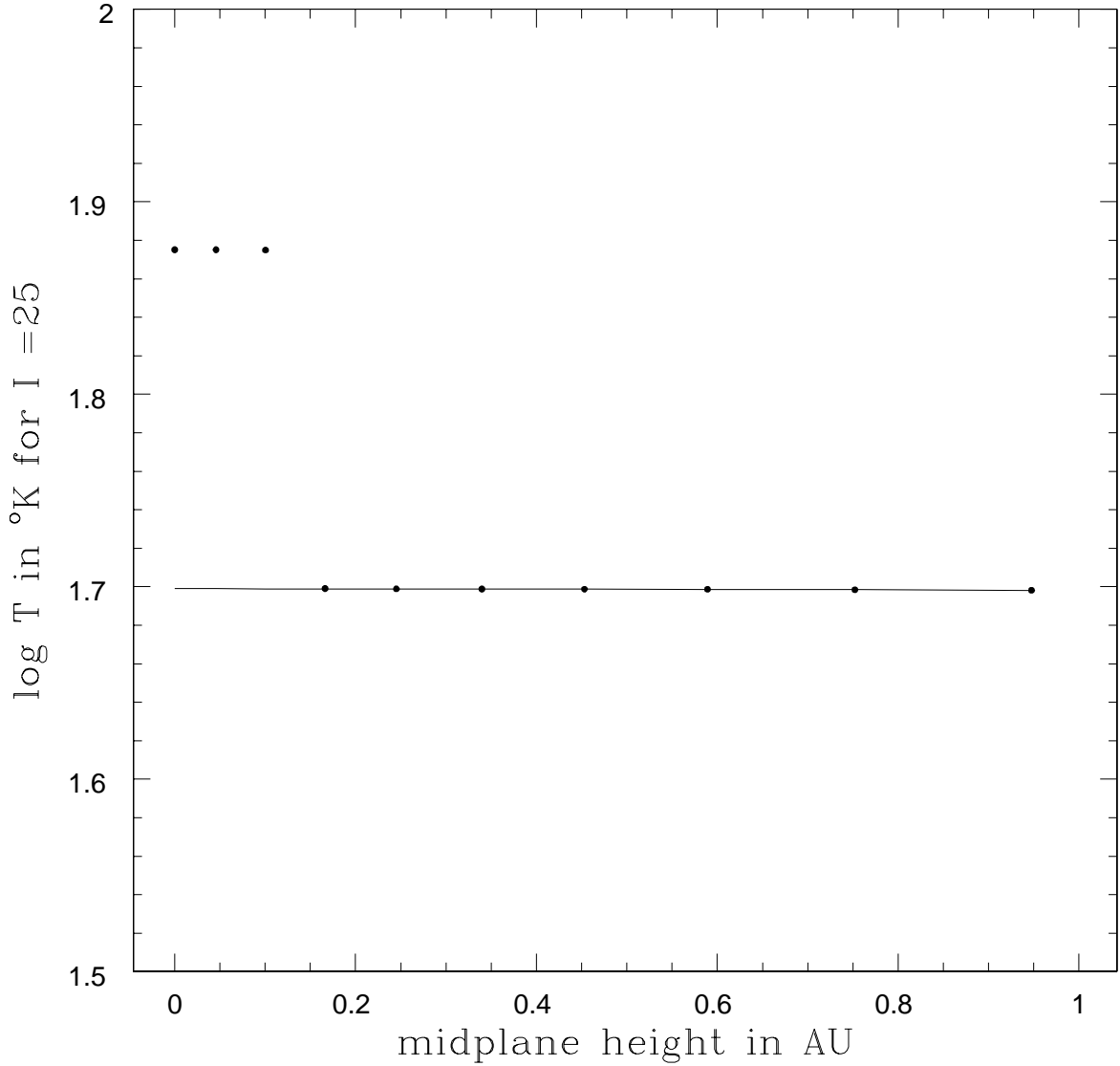}
\caption{Analytical steady-state vertical ($\theta$) temperature profile 
(solid line) and numerically-calculated temperatures (solid dots) for 
model T+50 after 0.19 yr of evolution at a radius of 7.87 AU. The initial 
50$\%$ temperature perturbation is evident between the disk midplane
and a height of 0.1 AU.}
\end{figure}

\clearpage

\begin{figure}
\vspace{-2.0in}
\plotone{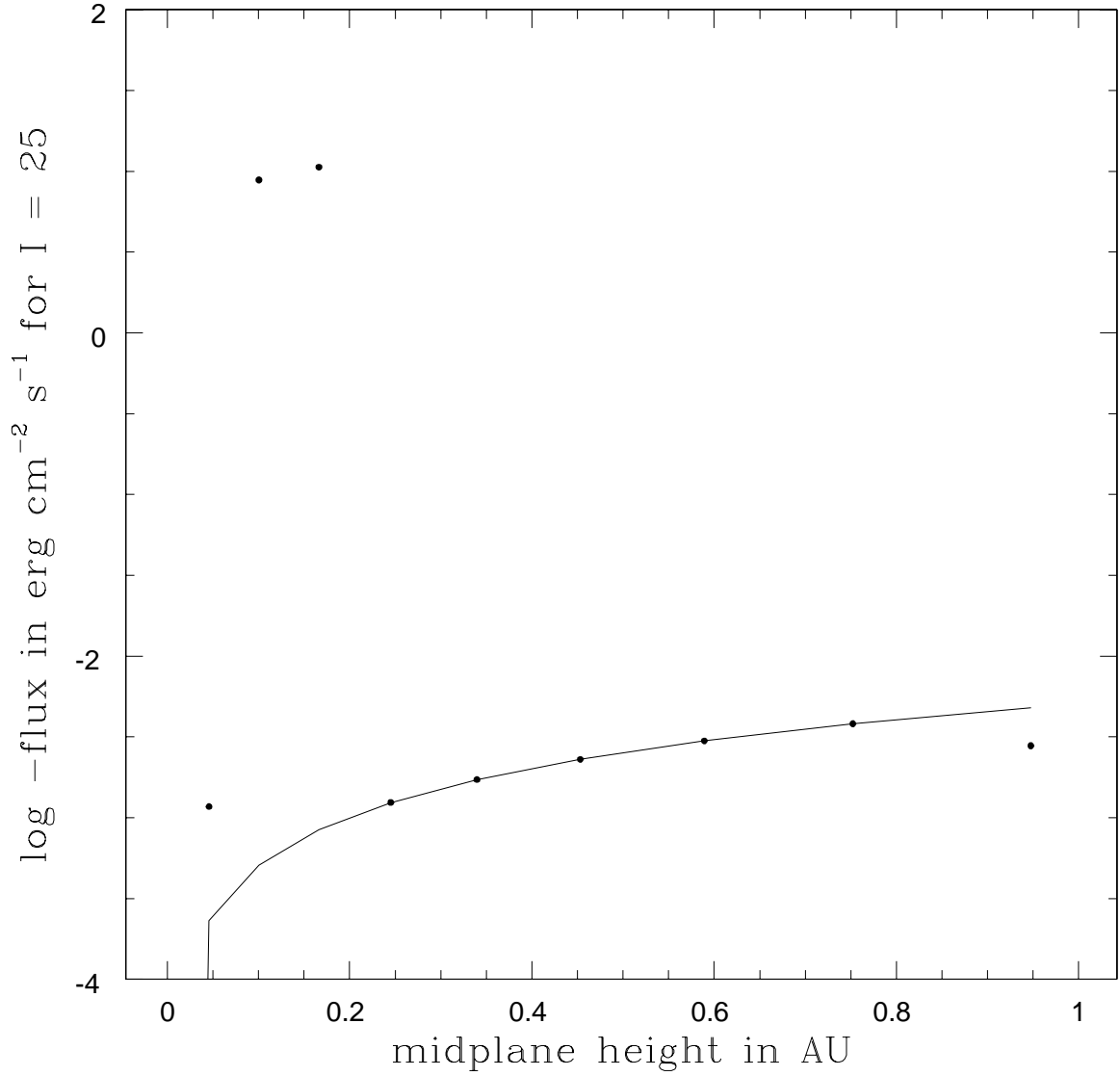}
\caption{Same as Figure 7, but for the radiative flux profile. 
The radiative flux falls to zero at the midplane.}
\end{figure}

\begin{figure}
\vspace{-2.0in}
\plotone{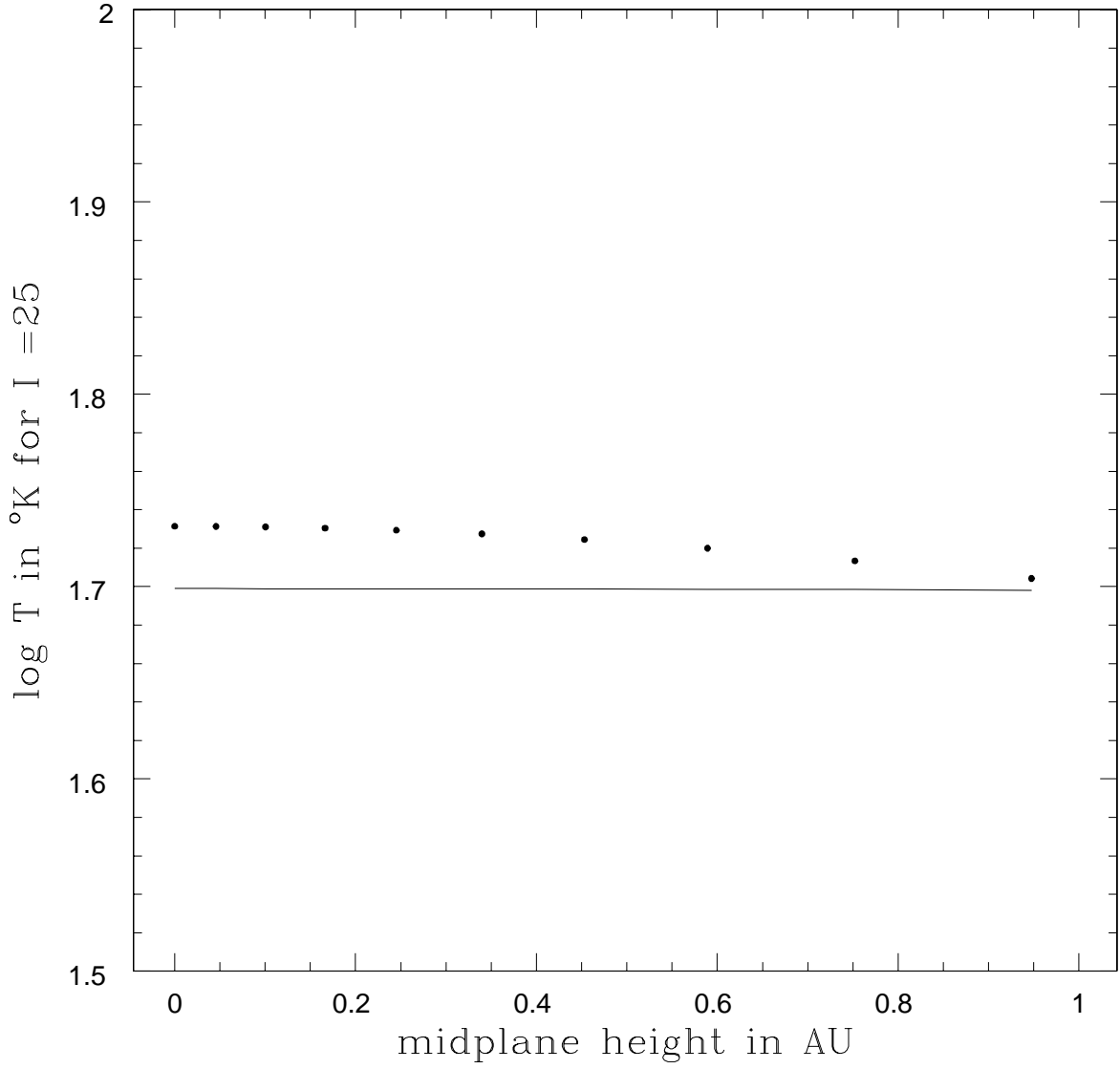}
\caption{Same as Figure 7, but after 0.019 Myr.}
\end{figure}

\clearpage

\begin{figure}
\vspace{-2.0in}
\plotone{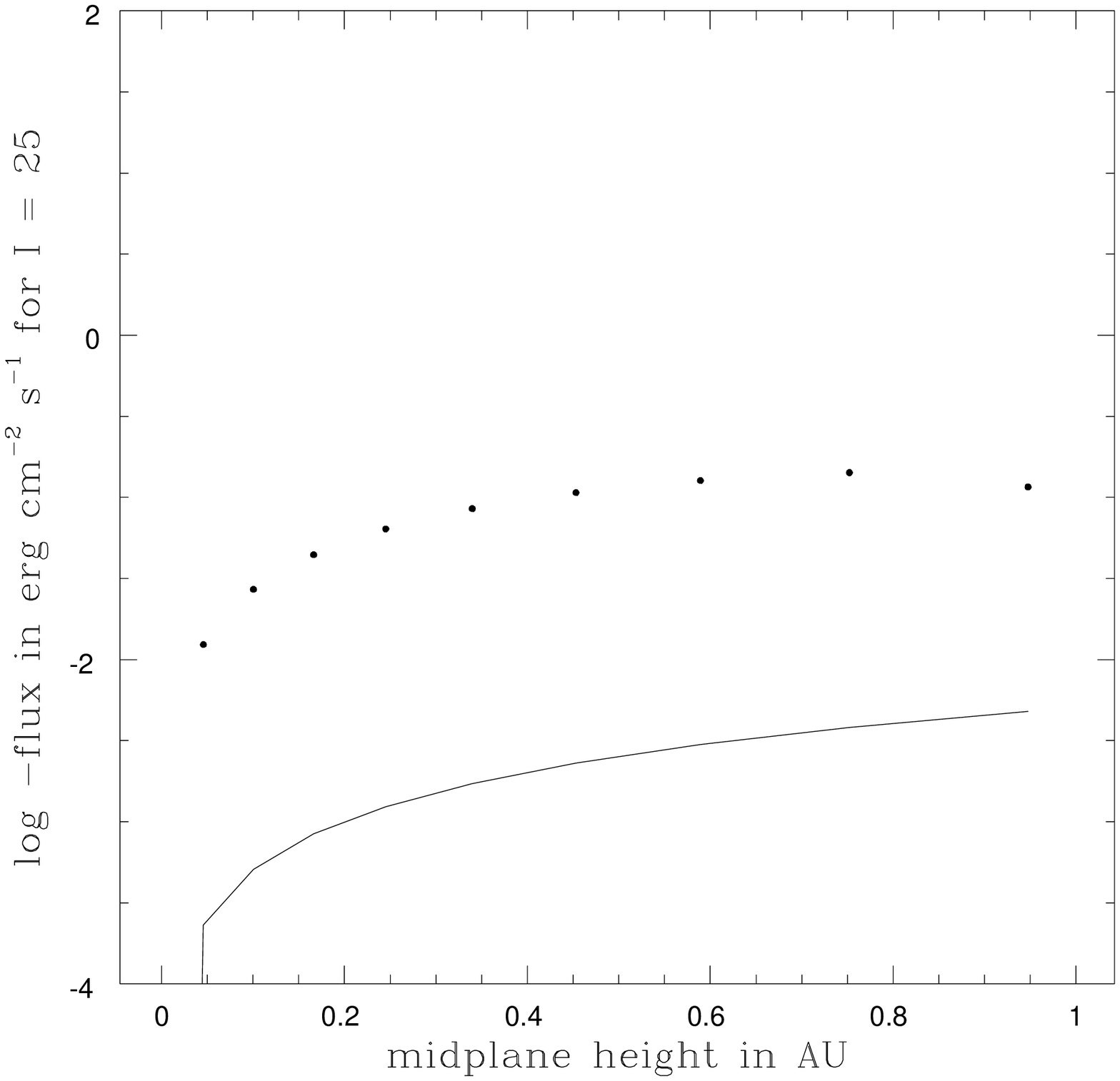}
\caption{Same as Figure 9, but for the radiative flux profile.}
\end{figure}

\begin{figure}
\vspace{-2.0in}
\plotone{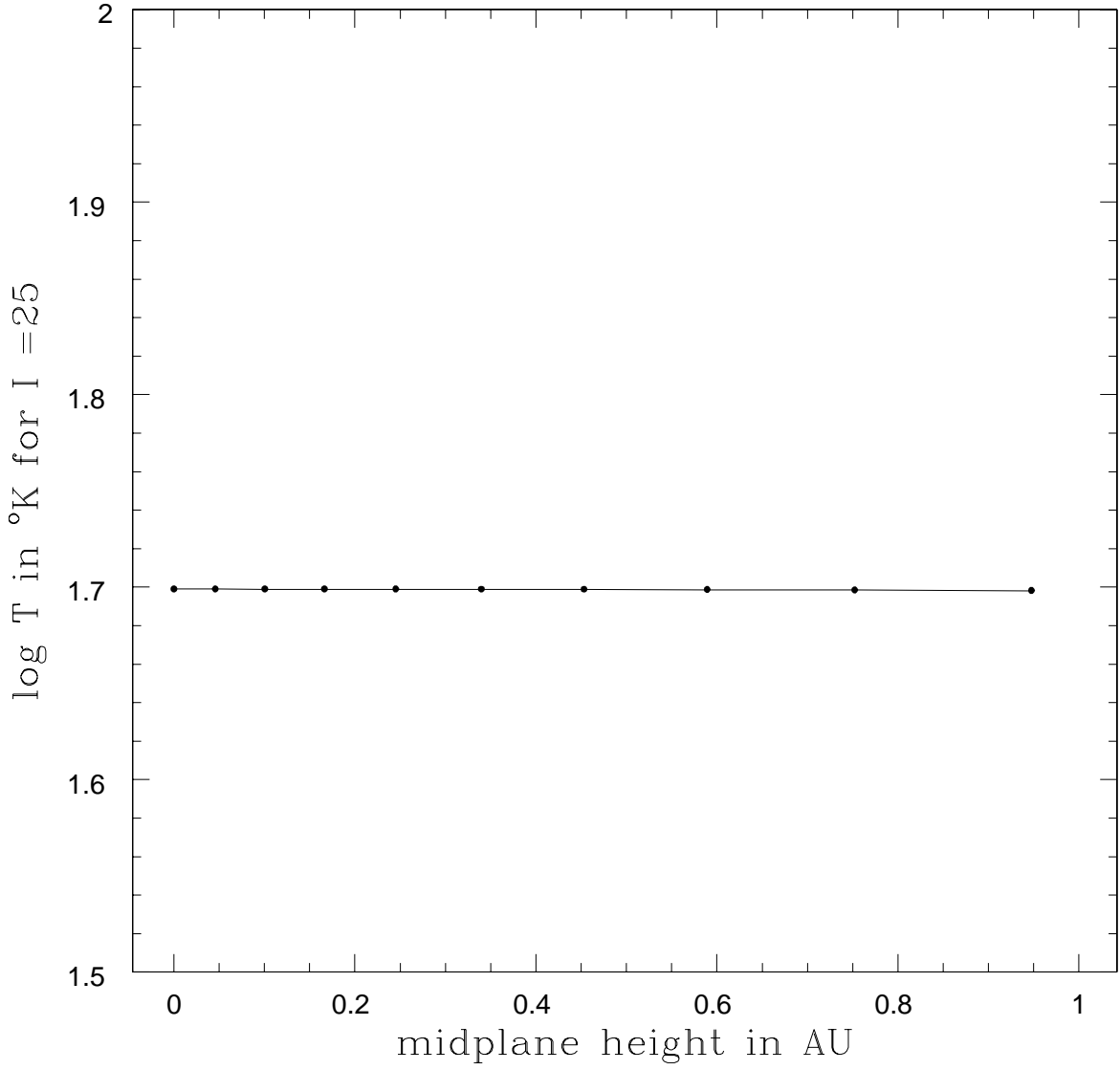}
\caption{Same as Figures 7 and 9, but after 1.9 Myr.}
\end{figure}

\clearpage

\begin{figure}
\vspace{-2.0in}
\plotone{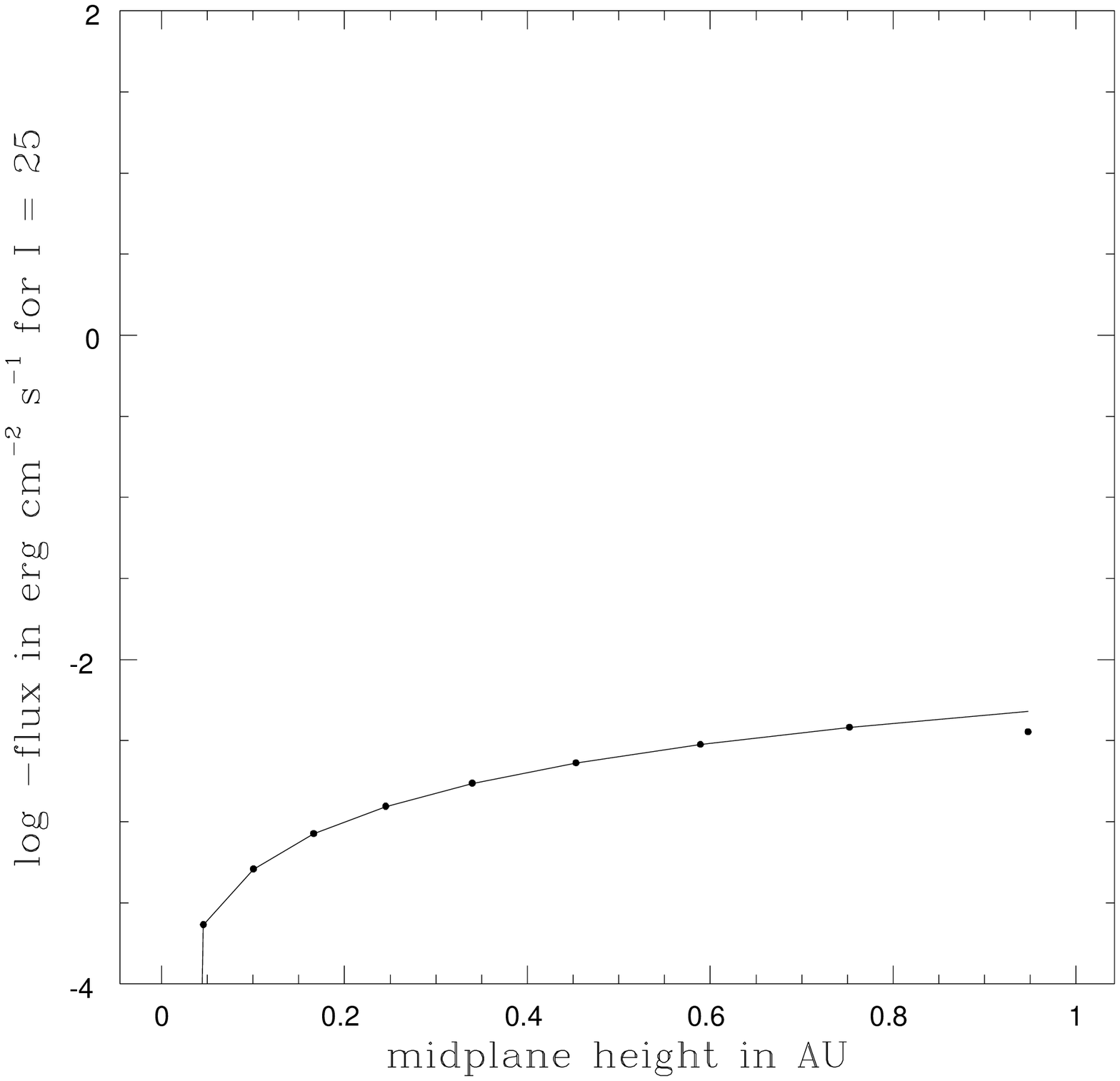}
\caption{Same as Figure 11, but for the radiative flux profile.}
\end{figure}

\end{document}